\documentclass[journal]{IEEEtran}
\ifCLASSINFOpdf
   \usepackage[pdftex]{graphicx}
   \DeclareGraphicsExtensions{.png}
\else
\fi
\usepackage[cmex10]{amsmath}
\usepackage{amsfonts}
\usepackage{algorithmic}
\usepackage[usenames]{color}

\graphicspath{{figures/}}

\begin{document}
\title{Sidelobe Canceling for Optimization of Reconfigurable Holographic Metamaterial Antenna}

\author{Mikala~C.~Johnson,~\IEEEmembership{Member,~IEEE,}
        Steven~L.~Brunton,~\IEEEmembership{Member,~IEEE}
				J.~Nathan~Kutz,~\IEEEmembership{Member,~IEEE}
        and~Nathan~B.~Kundtz,~\IEEEmembership{Member,~IEEE}% <-this % stops a space
\thanks{M. C. Johnson, S. L. Brunton, and J. N. Kutz are with the Department
of Applied Mathematics, University of Washington, Seattle,
WA, 98195 USA e-mail: mikalaj@uw.edu.}% <-this % stops a space
\thanks{M. C. Johnson and N. B. Kundtz are with Kymeta Corporation, Redmond,
WA, 98052 USA.}% <-this % stops a space
\thanks{Manuscript received ; revised .}}

\markboth{IEEE Transactions on Antennas and Propagation}%
{Johnson \MakeLowercase{\textit{et al.}}: Sidelobe Canceling}

\maketitle

\begin{abstract}
Accurate and efficient methods for beam-steering of holographic metamaterial antennas is of critical importance for enabling consumer usage of satellite data capacities.  We develop an optimization algorithm capable of performing adaptive, real-time control of antenna patterns while operating in dynamic environments.  Our method provides a first analysis of the antenna pattern optimization problem in the context of metamaterials and for the purpose of directing the central beam and significantly suppressing sidelobe levels.  The efficacy of the algorithm is demonstrated both on a computational model of the antenna and experimentally.  
Due to their exceptional portability, low-power consumption and lack of moving parts, metamaterial antennas are an attractive and viable technology when combined with proven software engineering strategies to optimize performance.  
%Due to its exceptional portability, low-power consumption and lack of moving parts, proven software engineering strategies optimizing performance make holographic metamaterial antennas an attractive and viable technology.

%
%Holographic metamaterial antennas are a promising technology to increase consumer usage of satellite capacity by providing an economical means of connectivity on mobile platforms.  Metamaterial antennas are built for portability - thin, lightweight, low-power consuming, and having no moving parts.  These antennas electronically beam-steer utilizing holographic principles; this means smart, adaptive software control algorithms are required for operation in dynamic environments. The prior literature has not directly addressed the issue of optimization methods to derive control patterns for a metamaterial antenna; this work does so.  It provides a first analysis of the pattern optimization problem, tying together two different viewpoints on the operating principles of the antenna, and then presenting one algorithm which is capable of significantly reducing specific sidelobes.  
%The efficacy of the algorithm is demonstrated both on a computational model of the system and experimentally.
\end{abstract}


\IEEEpeerreviewmaketitle

\section{Introduction}

\IEEEPARstart{T}{he} reconfigurable holographic metamaterial surface antenna (MSA) is an emerging technology for satellite communications.  The MSA is a low-power device that is flat, thin and lightweight.  Moreover, it achieves active electronic scanning without any mechanical moving parts.  All of these attributes make the MSA ideal for mobile satellite applications (automobiles, aircraft, trains, and ships).  However, in order to operate in dynamically changing environments, the antenna must be able to scan quickly, reliably, and without unacceptable levels of far-field radiation in undesired directions (sidelobes) .  It is therefore mandatory to suppress the production of sidelobes in a robust and adaptive fashion while still preserving a strong main beam to track a given satellite.  We develop an optimization algorithm for the far-field pattern of a MSA that explicitly addresses these issues.  Specifically, we demonstrate an algorithm for sidelobe suppression in software without resorting to non-adaptive hardware modifications. Further, we make explicit the connection between the disparate topics of reconfigurable metamaterial antennas, holography, and sidelobe control. 

Hardware developments for MSAs have recently undergone major innovations %IVantenna,
\cite{lipworth2013metamaterial,lim2004metamaterial,KymetaReceive}.  These advancements show that a metamaterial Ka-band antenna can be made manifest with existing materials and processes. Bi-directional high-speed internet connectivity was demonstrated by Kymeta Corporation with a metamaterial antenna in December 2013 \cite{KymetaTxRx}. However, the hardware antenna is only part of the system needed for this antenna to fulfill its industry-enabling promise. The antenna also must have smart controls to achieve optimal beam performance, being able to tailor, in a rapid and robust fashion, the radiation pattern of the antenna to achieve the desired characteristics that include a high peak gain, acceptable beamwidth, and sidelobe suppression.

Section \ref{sec:background} covers the background for classical antenna sidelobe cancellation and holographic antennas.  Section \ref{sec:systemmodel} describes the physical model and computational procedure used to predict the behavior of the antenna.  In section \ref{sec:theorydevel}, the theory connecting holography and the antenna field pattern is discussed. Section \ref{sec:algorithm} develops the algorithm used for sidelobe reduction. Finally, in section \ref{sec:results}, results demonstrating the efficacy of the algorithm on both the computational model of the system and on experimental hardware are demonstrated.  We then conclude the paper with discussion of the possibilities that our proposed control strategy provides. 

\begin{figure*}[!t]
\centerline{\includegraphics[width=7in]{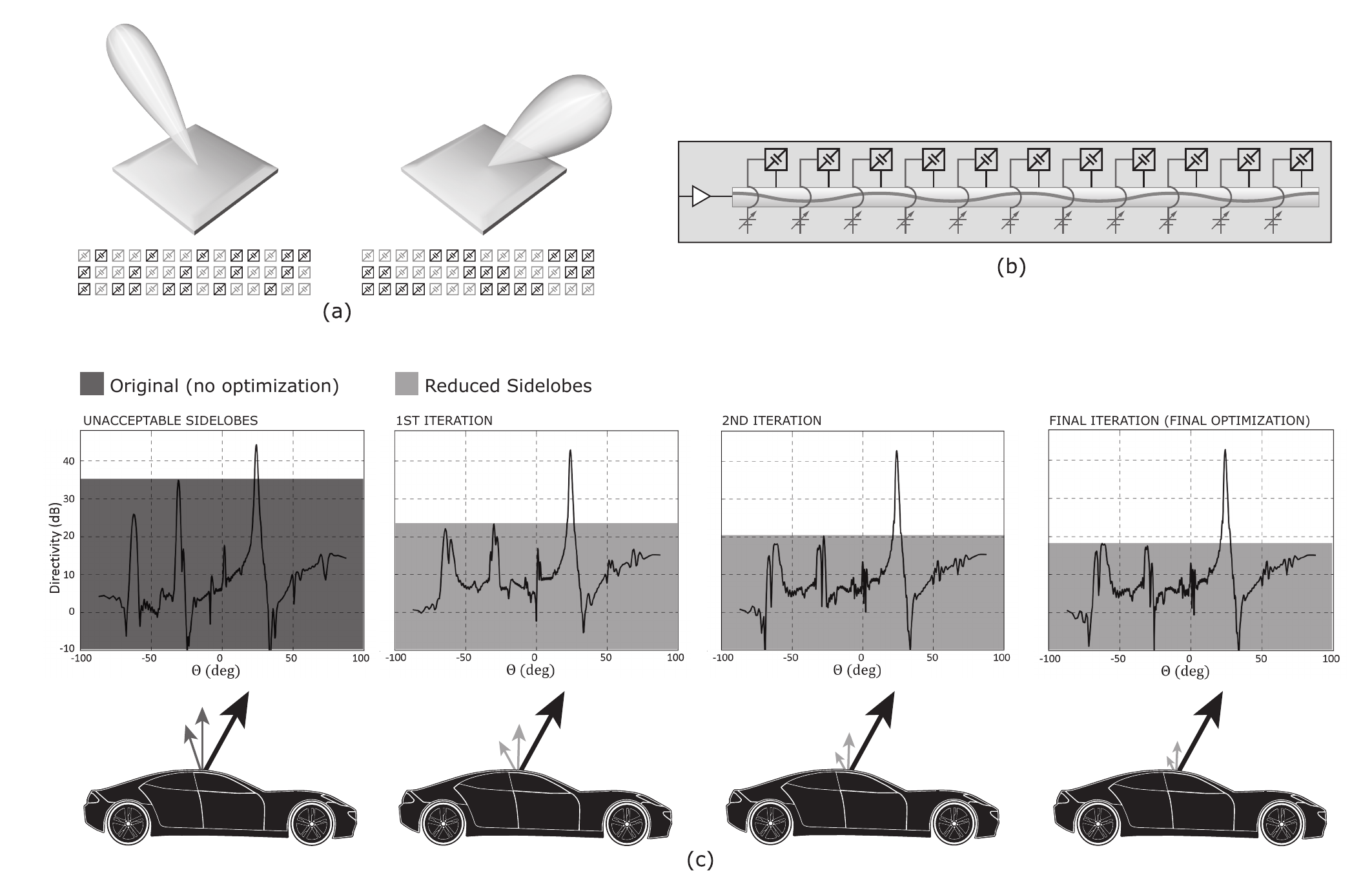}}%
\caption{(a) Holographic images on the surface of the array, when illuminated by an underlying feed wave, produce a coherent beam.  Different images, encoded in the array by the control applied to each meta-atom, produce beams in different directions. (b) The physics of the metamaterial antenna. The input carrier wave traverses the waveguide, and the meta-atom cells couple energy out of the feed.  Further, the cells produce simultaneous amplitude and phase-shift as a function of their control. (c) (top row) Simulated far-field results of the algorithm iteratively targeting different sidelobes while continuously creating a main beam in the direction of $25.7^{\circ}$ $\theta$. (bottom row) Representation of the simulated results in a mobile application.}
\label{fig:FancyImage}
\end{figure*}

\section{Background} \label{sec:background}

In this section, we give an overview of sidelobe cancellation along with an introduction to holographic antennas. 

\subsection{Sidelobe cancellation in arrays}

Sidelobe cancellation has a rich history in the academic literature as well as in practice. Several patents were issued for device configurations for sidelobe cancellation in the mid-1960's \cite{w1965intermediate,howard1969sidelobe,durboraw1983clutter,tsujimoto1994sidelobe}.  Unlike the operating principles of the metamaterial antenna, these patents dictate the use of spatially separate arrays for receiving the signal and for detecting the noise signals. During the processing of the signal, the interference signal is subtracted from the reference, and thus the unwanted information received from sidelobes are suppressed.

The mathematics of this interference-signal-subtraction approach was formalized in the Generalized Sidelobe Canceler that was published in 1975 by Widrow et al. as an application of a broadly-useful adaptive noise canceling algorithm \cite{widrow1975adaptive}. The algorithm dictates how to optimally tune the weights during signal processing of a receiving beamformer to remove the interference signal and minimize the loss of the desired signal. This noise-canceling approach to sidelobe reduction is demonstrated on a phased-array antenna in \cite{widrow1975adaptive} and many works that follow.  The generalized sidelobe canceler and its extensions assume that the signal from each antenna comprising the array is separable, which is not the case in the holographic antenna \cite{baird1976adaptive,frost1978multilevel,griffiths1982alternative,Nordholm2000,herbordt2001efficient,oak2005calibration}.     
%Assuming that the reference signal contains both noise and the desired signal and that the second signal contains primarily just the noise (an omnidirectional antenna), one subtracts the noise signal directly from reference signal.  If, of course, the secondary signal contains the reference signal as well, the desired signal will be attenuated.  The way around this issue is to "block the main signal from the calculation of the background in the second signal.  In this algorithm, it is assumed that one knows the desired frequency or direction of arrival of the desired signal, then the rest of the spectrum of the received signal is subtracted out in proportion to its contribution to the signal.
%More theoretical analysis and extensions to the generalized sidelobe canceler were forthcoming after Widrow.  Other extensions and examinations of this method which added complexity and increased robustness or efficiency, but did not represent significant departures in approach or application, include increasing computational efficiency by manipulation in the frequency domain These papers included analysis of the robustness of the algorithm to errors in pointing (amplitude and/or phase errors in the beamforming weights). 
The current work develops a sidelobe cancellation algorithm that works under the same principle as the noise-canceling algorithm of Widrow et al., but with important and non-intuitive extensions needed for the MSA.  This work also constructs a low-dimensional parameterization of the control pattern for this type of antenna.

\subsection{Holographic antennas}
A microwave holographic antenna was first demonstrated by Checcacci, Russo, and Scheggi in 1970 \cite{checcacci1970holographic}. %The holographic interference pattern at microwave frequencies was first recorded at Bell Labs in 1950, but was only used for imaging rather than for attempting to reproduce the three-dimensional image, much less form a microwave-frequency beam \cite{kock1975engineering}.
Particularly interesting in this work is that their first interest beyond creating a beam-forming antenna was to reduce or eliminate the \lq\lq zeroth order\rq\rq \hspace{1pt} aberrations in the holographic image.  
They do not call these aberrations sidelobes, although that is what they are. 
%They do not give these aberrations the name sidelobes, although that is what they are.  
They understood these aberrations to be errors in the holographic recording and addressed them accordingly, varying the approximation (pixelation and quantization) of the holographic recording to reduce this zeroth order.

Holographic antennas have been under development for varying frequencies of operations since Checcacci's time. The implementations include static artificial impedance surfaces \cite{fong2010scalar} as well as reconfigurable holographic antennas, e.g. \cite{taylor2003reconfigurable}. This literature typically focuses on the hardware devices that can accomplish the recording and illumination of a microwave hologram. These works do not consider optimization of the beam in software, but instead focus on material developments to achieve more accurate reproduction of the holographic interference image with techniques like increasing pixel density or increasing the range of phase control.
We specifically address optimization of the reproduced holographic image in software control despite pixelation and control quantization. 
%This work examines, analytically and empirically, the parameter space of control pattern generation for a holographic metamaterial antenna.
%One descriptive explanation of the metamaterial antenna which is not published elsewhere is to begin with the examination of the holographic antenna concept. We are all familiar with optical holograms.  These are pieces of special paper on which is recorded the image of a three-dimensional object 
%The metamaterial antenna with its densely packed cells operate using holographic principles to form a coherent, directed beam. Holography, in this context,  means exciting an image of the desired far-field pattern on the surface of the array.
%
%The holography is also affected by the discrete nature of the meta atoms.  Thus, errors are induced in localization of the current distribution on the surface of the array.  Quantization errors are introduced in the controls themselves.  A very discrete set of states is achievable 

\section{System Model} \label{sec:systemmodel}

%In this section, we briefly describe the holographic metamaterial antenna system physics that are most relevant to the problem of controlling the antenna, and we describe the computational model of the antenna that was used as a means of studying the efficacy of the proposed control strategy.
%Assuming idealities of the antenna like a continuous, smoothly varying distribution of current on the plane of the antenna and trivial coupling between elements, the principles of holography can be used to prescribe exactly the state to which one wants the antenna to be controlled to achieve a perfect pencil beam in the far-field. However, the approximations which must be made and the quantization which must take place to achieve a real device along with the uncertainties in the produced piece of hardware yield significant challenges to high-fidelity control.
%NEED A TRANSITION HERE OR THIS PARAGRAPH DOESN"'T BELONG HERE

The MSA technology architecture is composed of thousands of discrete meta-atoms (equivalently, unit cells, resonant cells, or resonators) packed closely together in a rectangular array and fed by a propagating feed/carrier wave beneath the elements. This structure leads to several characteristics that must be considered when approaching optimization of the radiation pattern.

First, a metamaterial antenna capable of closing a link with a satellite must have enough surface area to attain high enough gain.  To achieve this basic performance requirement, the antenna is composed of more than 10,000 continuously and individually controlled unit cells. The control of each cell varies the amplitude of scatter of the cell from a minimally excited state to a maximally excited state. It is impractical and infeasible to try random controls, e.g. a genetic algorithm, in search of good control patterns. There is, however, a small manifold within this massive parameter space that yields a coherent beam. The explication of the algorithm in section \ref{sec:algorithm} specifically points out this useful (and unsurprising) low-dimensional parameterization that permits defining excellent control patterns quickly.

Apart from the large number of cells, each cell responds nonlinearly in both phase and amplitude to its control. The cells do not display \textit{independent} amplitude and phase modulation as a phased array would, but instead are resonant; the amplitude and phase shift happen simultaneously as the control is changed. The resonance behavior, particularly the phase shift from the underlying carrier wave, is especially sensitive to manufacturing tolerances.  Naturally, we seek an optimization algorithm that is flexible and robust to such real-world tolerances. The need for flexibility in the face of uncertainty points us toward an iterative algorithm where the antenna can learn about itself to improve its control pattern intelligently with feedback.

The coupling between the elements and the coupling between all the elements simultaneously with the feed wave in the waveguide further complicates control. All elements are \textit{simultaneously} slightly changing the feed wave based upon the applied controls. Upstream elements couple energy out such that downstream elements have less energy exciting them. Also, downstream elements incite a small backward propagating wave that is typically evanescent, but the elements are so closely spaced that even an evanescent wave may affect near-neighboring cells. Additionally, the coupling to the feed wave of each cell is dependent upon its control value as one hopes and expects, but this complex simultaneous response is very important in accurately defining the correct holographic image.
%The important physical effects creating phase variations are those due to the traveling wave and, additively, the phase shift which occurs through the meta-atom, particularly when it is highly excited. The largest amplitude variations evidenced by the individual cells are due to the decaying power down the waveguide and the control of each of the cells. 

%The corporate feed also means that all elements are \textit{simultaneously} interacting with and slightly changing the feed wave based upon the controls themselves: upstream elements couple energy out so that downstream elements have less energy exciting them. Downstream elements incite a small backward propagating wave that is typically evanescent, but the elements are closely spaced so that even an evanescent wave may effect near neighboring cells. The coupling to the feed wave of each cell is dependent upon its control value. This complex simultaneous response will be demonstrated to be a very important aspect to an accurate definition of the holographic image. 
A schematic description of these complex coupled physics of the array is shown in Figure \ref{fig:FancyImage}(b). This image shows the traveling feed wave and the cells coupled to the waveguide.  It also indicates the phase-shift from the underlying carrier phase due to the resonance of the cells. This structure in which all cells are coupled to the waveguide as well as to each other and in which the phase shift is dependent on the control poses significant challenges for control and optimization.

A discrete dipole approximation-based model (DDA model) of the metamaterial antenna captures all the effects mentioned above and has been demonstrated to more accurately predict the far-field radiation pattern of the antenna than other models typically used in metamaterial modeling, while being fast enough for rapid iteration for optimization \cite{johnson2014}. Thus, in this work we use a DDA model of the antenna to demonstrate our optimization scheme. Refer to \cite{johnson2014} for details on the properties and parameters of this model.  In this work, the planar MSA is simplified to a one-dimensional array; a single row of meta-atom resonators. Note that in the results given in section \ref{sec:results}, the DDA model actually incorporates a hardware tapering that causes every cell in the strip to possess different control response behaviors; the DDA is particularly well-suited to modeling of a metamaterial when the unit cells are not identical. 

\section{Theoretical Developments} \label{sec:theorydevel}

Two perspectives are given on the beamforming of the MSA.  First, we discuss the beamforming from the phased array perspective based upon an array factor, i.e. the summation of the fields radiating from each individual cell.  Next, we explore beamforming from a holography viewpoint and highlight where the two interpretations connect.  Particular attention is given when the holographic explanation sheds light on important considerations for achieving control of the antenna pattern.
%In particular 
%The image of a spherical wave from a satellite on the surface of a small metamaterial antenna is a plane wave with a single spatial frequency. When the discrete elements and finite size of the metamaterial antenna fails to replicate exactly the image (in transmit or receive), sidelobes manifest; energy is scattered into or received from unintended spatial directions. Control pattern generation for the metamaterial antenna focuses on predicting these errors and minimizing them for lower sidelobes and increased gain resulting in higher signal to noise ratio, higher EIRP (equivalent isotropically radiated power) when discussing transmit or higher G/T (Gain over temperature) in antenna terminologies.

\subsection{Array theory development}

Following is a short derivation of the basic control theory for a metamaterial antenna based upon familiar mathematical concepts from phased array theory.  The far-field of an array of point sources is given by the array factor:
\begin{equation}
AF(\theta_0,\phi_0) = \sum^{N}_{n=1} A_n \exp(-i\mathbf{k}_f(\theta_0,\phi_0) \cdot \mathbf{r}_n),
\label{eq:arrayfactor}
\end{equation}
where $\mathbf{k}_f(\theta_0,\phi_0) $ is the desired directional complex propagation vector in free space, and $\mathbf{r}_n$ is a coordinate of a radiator on the surface of the antenna.  

$A_n$ are the complex weights of the individual antennas comprising the array. In a classical active phased array, both the amplitude and phase of these coefficients can be selected to achieve a desired far-field pattern. In the metamaterial antenna, these properties cannot be independently selected.  However, since phase shifters are not being used, the power needed to steer the antenna is 1/100th or better than that of a phased array.  The phase of the complex weight is largely the result of two effects, namely the traveling wave phase and the additional phase shift from the base traveling wave phase induced by the resonant properties of the cells.  Thus the phases of the $A_n$ are constrained variations around the traveling wave phase. The achievable amplitude range between \lq\lq off\rq\rq \hspace{1pt} (minimally excited cell) and \lq\lq on\rq\rq \hspace{1pt} (maximally excited cell) is limited to values as small as 1:3 or up to about 1:15. We can think of the selection of the $A_n$ as being a severely constrained version of a phased array, or we can think of them as a holographic recording mechanism as described in the next section.

As is well-noted, the far-field is related to the array weight distribution by a Discrete Fourier Transform.  We seek to create a delta function of the energy in a particular direction in the far-field, aligning the phase distribution attained from the traveling wave with the desired radiating phase distribution:
\begin{equation}
\sum_{n=1}^{N} \mathbf{E}_n (\mathbf{r}_n; \mathbf{k}_s) \propto \delta (\mathbf{k}_f(\theta_0,\phi_0) - \mathbf{k}_s),
\label{eq:sumfields}
\end{equation}
where $\mathbf{k}_s$ is the complex propagation vector of the reference traveling wave, $\mathbf{E}_n (\mathbf{r}_n; \mathbf{k}_s)$ is the electric field generated by a radiator at location $\mathbf{r}_n$, and $\delta$ is the Dirac delta function.

If the array were infinite and a continuous current distribution, the Fourier analysis indicates that the distribution of the weights should then be:
\begin{equation}
A(\mathbf{r}) = \exp(-i \mathbf{k}_f(\theta_0,\phi_0) \cdot \mathbf{r}) \exp(i \mathbf{k}_s \cdot \mathbf{r}).
\label{eq:fourierweights}
\end{equation}
%Obviously, we have to truncate and discretize based on the location of the resonators. 
Note that these ideal weights are complex, and we must apply real controls to approximately achieve these desired weights. Additionally, given the assumed phase distribution on the array $ \exp(i \mathbf{k}_f \cdot \mathbf{r})$, some cells will be displaying exactly the correct phase indicated by equation \ref{eq:fourierweights} while most of them will be radiating a phase which is incorrect, at the extreme case, being $180^{\circ}$ out of phase. A control strategy may then be to allow cells at the correct phase to radiate strongly and to disallow cells at the incorrect phases to radiate. This strategy is parameterized mathematically by:

\begin{equation}
m(\mathbf{r}_n;\theta_0)= \frac{\Re(A_n) + 1}{2}.
\label{eq:control}
\end{equation}

This real-valued control is a shifted and scaled cosine wave, $\cos((\mathbf{k}_s-\mathbf{k}_f(\theta_0,\phi_0)) \cdot \mathbf{r})$. A depiction of sinusoidal control patterns producing different beams is shown in Figure \ref{fig:FancyImage} (a). Control patterns of different length period produce different beams as indicated by equation \ref{eq:control}. Further, this control of each individual cell is, in fact, a sampling of this cosine wave. The well-studied relationships of the DFT and the Fourier Transform hold for this discretized antenna array case indicating that sampling the waveform induces sidelobes due to aliasing.   In fact, harmonic sidelobes of the form $\cos(m\times(\mathbf{k}_s-\mathbf{k}_f(\theta_0,\phi_0)) \cdot \mathbf{r})$ with $m\in \mathbb{Z}^+$ are the largest lobes in measurement and simulation.  Gibbs ringing due to a finite array-size is also evident in far-field pattern.  Sidelobes due to these effects are always going to be manifest, but the largest from either effect can be significantly reduced with the algorithm proposed in section \ref{sec:algorithm}.

\subsection{Holographic theory development}
Alternatively, one can consider the control problem from the holographic perspective and derive the same control strategy.  This second perspective is particularly helpful for identifying additional limitations of the above noted control strategy beyond the finite array size and the pixelation/sampling of the cosine waveform.  The theory of holography is to encode on a surface (in this case, the antenna) an image such that, when the surface is illuminated by a specified reference wave, the viewer sees a complete version of the originally recorded 3D image. Refer to Figure \ref{fig:FancyImage}(a).

In the case of a MSA, consider the desired 3D image to be a spherical wave emitted by (or converging upon) a point source in the far-field (the satellite).  This propagating wavefront appears as a plane-wave on the flat surface of the antenna on earth. The illuminating reference wave is the wave that lights up the meta-atom resonators: the propagating electric-field in the waveguide.

\begin{eqnarray}
\Psi_{obj}(\mathbf{r};\theta_0,\phi_0) &=& \exp(-i \mathbf{k}_f(\theta_0,\phi_0) \cdot \mathbf{r})
\label{eq:waveout} \\
\Psi_{ref}(\mathbf{r}) &\approx& \exp(-i \mathbf{k}_s \cdot \mathbf{r} ), 
\label{eq:wavein}
\end{eqnarray}
where $\Psi_{obj}$ is the desired far-field wave and $\Psi_{ref}$ is the illuminating wave in the waveguide.  In particular, we are interested in coordinates $\mathbf{r}$ that are on the surface of the antenna.
%, $\mathbf{k}_f(\theta_0,\phi_0) $ is the desired directional complex propagation vector in free space, $\mathbf{k}_s$ is the complex propagation vector of the reference wave

The \textit{transmittance} represents the \lq\lq picture\rq\rq \hspace{1pt} on the surface derived from the simultaneous presence (or summation) of the two waves.
\begin{eqnarray}
T \propto |\Psi_{obj} + \Psi_{ref}|^2 &=& |\Psi_{obj}|^2 + \Psi_{obj}^*\Psi_{ref} + \notag \\
&&\Psi_{obj}\Psi_{ref}^* + |\Psi_{ref}|^2.
\label{eq:transmittance}
\end{eqnarray}
$T$ is proportional to the right hand side of this equation by a scalar (potentially complex value) indicating that small amounts of energy may be lost or dispersed by other mechanisms.

If this transmittance picture is \lq\lq recorded\rq\rq \hspace{1pt} in some way, then lit up by the original reference wave, we obtain:
\begin{eqnarray}
T\Psi_{ref} &\propto& |\Psi_{obj}|^2\Psi_{ref} + |\Psi_{ref}|^2\Psi_{ref} + \notag \\
&&\Psi_{obj}^*\Psi_{ref}^2+\Psi_{obj}|\Psi_{ref}|^2.
\label{eq:re-illumination}
\end{eqnarray}
The four terms appearing in the imaging of the hologram represent different physical terms.  

The first term $ |\Psi_{obj}|^2\Psi_{ref} $ has amplitude proportional to the object beam and points in the propagating direction of the reference wave.  This image does not show up in the hologram; its image and energy are transferred through with the reference wave. In the holographic antenna case, the energy in this portion of the image is transmitted down the waveguide and attenuated at the end. The second term is similar to the first in that it is in the direction of the reference wave and does not produce an image.
The third term $\Psi_{obj}^*\Psi_{ref}^2$ produces an image of the conjugate to the desired beam. This image for the holographic antenna is typically not in the visible spectrum of the array.   The fourth term $\Psi_{obj}|\Psi_{ref}|^2$ is the term of most interest.  It has an amplitude proportional to the input wave, and it points in the direction of the desired object beam. 
%To get maximal energy transfer in this term, the interference image needs to be recorded correctly and the reference beam which is used in the calculation of the transmittance must be exactly the conjugate of the illuminating reference wave.  
We will refer often to the wave interference as the portion of the transmittance which is desirable:
\begin{equation}
\Psi_{intf} =: \Psi_{obj}\Psi_{ref}^*.
\label{eq:waveinterference}
\end{equation}

To get maximal energy transfer in this term, the interference image needs to be recorded correctly and the reference beam in the interference calculation needs to exactly match the illuminating wave $\Psi_{ref}$.  Any errors in this imaging process will cause the object beam to not be exactly replicated.  

For the metamaterial antenna considered here, there are a number of deviations from these ideal equations that will occur. The finite size of the array limits the resolution of the image in the far-field and reduces the field of view (the scan range). The pixelation of the array also limits the resolution of the far-field and the ability to scan as far. Note that the limited resolution of the holographic image can be understood in the context of the Discrete Fourier Transform as described above.  
%Discretization and the finite size of the array induces sinc-type behavior in the far-field pattern.  That is, immediately, a finite and discretized array manifests sidelobes.

The other difficulty in holography, which is not obvious from the array theory analysis, is that the reference wave in the wave interference and the illuminating wave need be exactly the same or there will be aberrations in the resulting holographic image.  However, the application of the controls to the individual meta-atoms to record a desired interference image actually affects the illuminating reference wave enough to cause errors. 
%The control pattern of the array is our means of recording, \lq\lq taking a picture of\rq\rq, the desired wave interference image.  Loosely speaking the control selects the amount of the traveling wave (reference wave) which is permitted to leak into free space.

The basic wave interference in one-dimensional to scan to a given angle $\theta_0$ is given by:
\begin{equation}
\Psi_{intf}(\mathbf{r};\theta_0) =: \exp(-i\mathbf{k}_f(\theta_0) \cdot \mathbf{r}) \exp(i\mathbf{k}_s \cdot \mathbf{r})
\label{eq:intf}
\end{equation}
Note that this equation assumes that the illumination is from an ideal traveling wave within the waveguide; it assumes that the control pattern (or anything else) does not change this propagation and that there is no phase shift between the underlying traveling wave and the radiated energy due to the resonant behavior of the cells.  Both of these assumptions are false but provide a useful beginning point for discussion.  In fact, the above assumption is accurate enough 
%\lq\lq true enough\rq\rq \hspace{1pt} 
to ensure that the main beam points in the intended direction.  The problem is the resolution of the rest of the pattern, i.e. the unintended sidelobes induced by the inaccuracy.

This wave interference equation derives exactly the same weighting and control strategy as was found in the array factor analysis equations \ref{eq:fourierweights} and \ref{eq:control}. But this derivation gives additional understanding of the potential for sidelobes in the resulting far-field as coming from a mismatch of the reference wave used for calculation of the control and the reference wave that actually illuminates the array due to the application of the controls themselves.

Since the application of the control itself induces complex errors in the illuminating reference wave which produces sidelobes, an iterative algorithm must be used to find optimal control patterns. Sidelobes are also produced by the finite array, the discretization of the cells, the quantization of the control states, and the uncertain characteristics of the array due to tolerances in real materials. Thus, we also require a general approach that allows us to target any sidelobe no matter how it is induced. In the next section, we present a generalized and iterative algorithm for sidelobe cancellation.  

\section{Algorithm} \label{sec:algorithm}

\begin{figure}[!t]
\centerline{\includegraphics[width=3.5in]{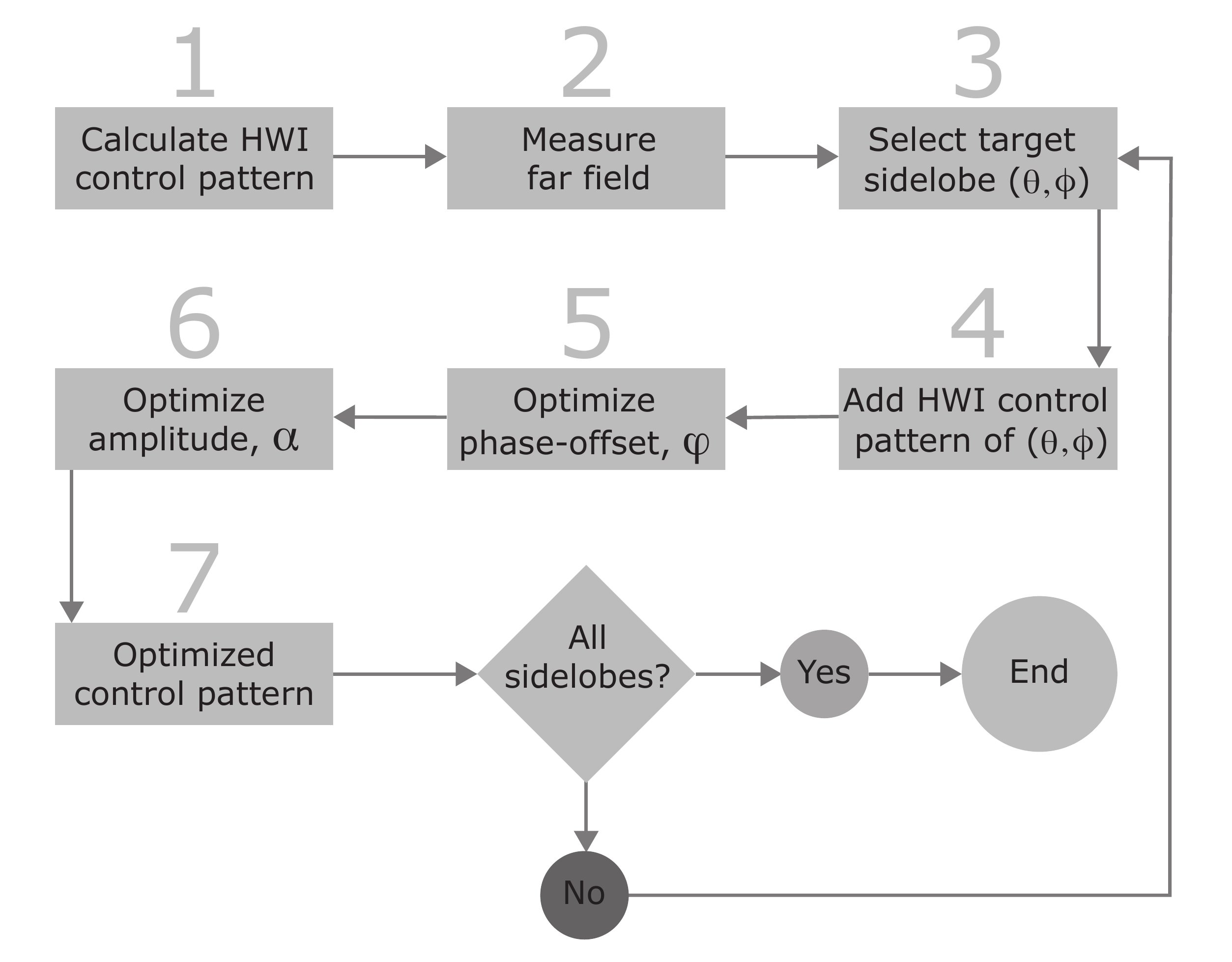}}%
\caption{Flowchart of sidelobe reduction algorithm. HWI stands for Holographic Wave Interference, see equation \ref{eq:waveinterference}}
\label{fig:Algorithm}
\end{figure}

\begin{figure*}[!t]
\centerline{\includegraphics[width=3.5in]{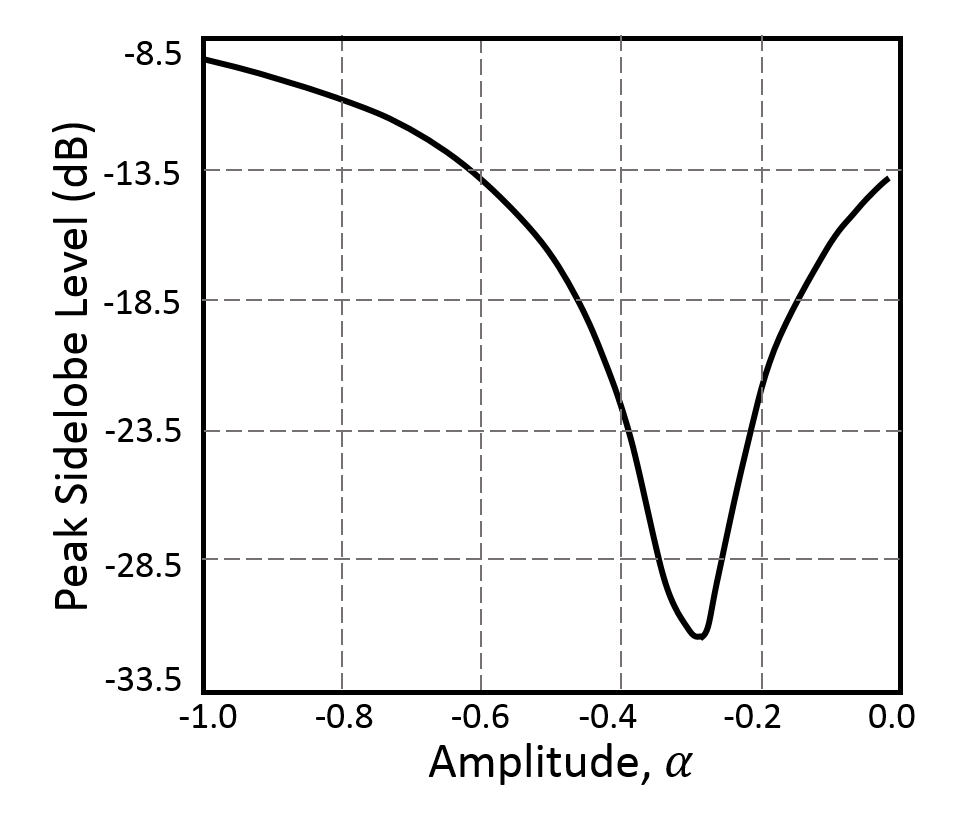}\label{fig:amplitudesmooth}
\hspace{1pt}
\includegraphics[width=3.5in]{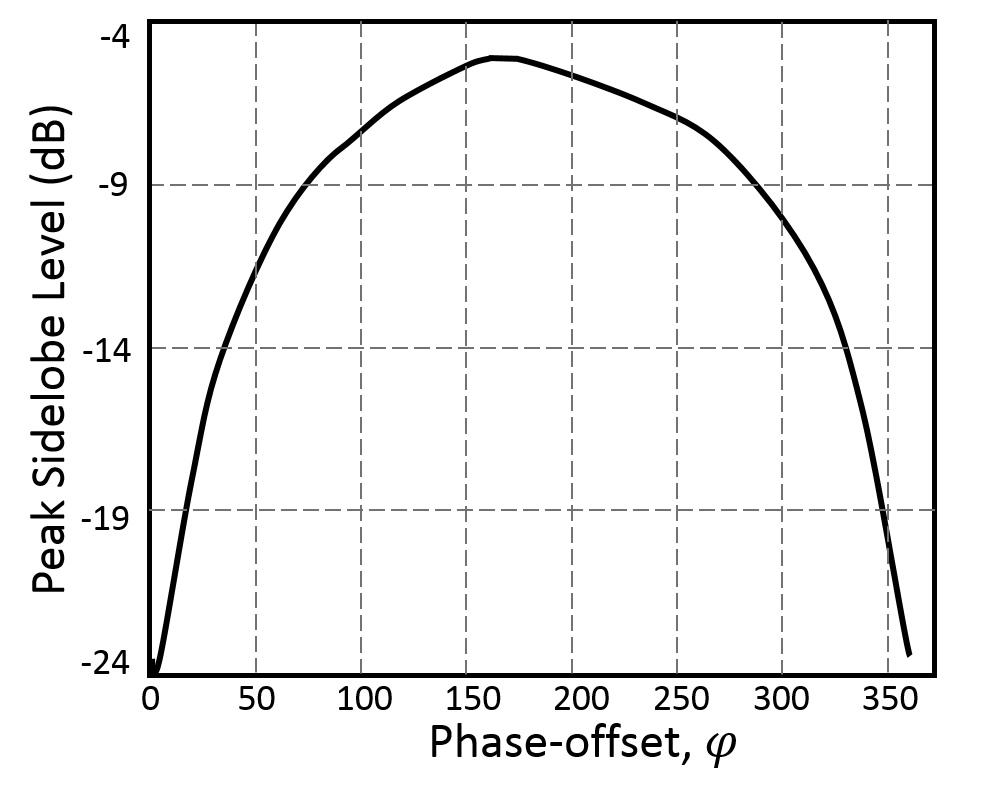}\label{fig:phasesmooth}}
\caption{Sidelobe level as a function of (left) amplitude of auxiliary pattern and (right) phase-offset of auxiliary pattern for the example of an array pointing at $-30^{\circ}$ and targeting a sidelobe at $+4^{\circ}$.}
\label{fig:smooth}
\end{figure*}

In this section, we develop our algorithm for finding optimal control patterns for the MSA, which reduces targeted sidelobes. 
%This algorithm implements the principles of noise-canceling which were given by Widrow. However, this manifestation of noise-canceling is specific to a holographic antenna, an antenna in which the signal from every antenna composing the array cannot be separated out, 
%% MIKALA:  this next sentence doesn't make sense
%and an antenna in which much more dense as it is resolving and encoding a wave interference pattern.
The sidelobe targeting algorithm starts from the control strategy that was derived in section \ref{sec:theorydevel} and is given in equation \ref{eq:control} to be a discretized cosine wave, step 1 in Figure \ref{fig:Algorithm}. 

First, note that the control defined by equation \ref{eq:control} dictates the controls across the whole array, all $10,000+$ cells, as a function of very few parameters, namely the desired pointing vector $\mathbf{k}_f(\theta_0,\phi_0)$. That is, the wave interference equation gives a low-dimensional manifold that yields coherent beams. To address sidelobes we slowly enlarge this parametric space by adding a second waveform, found on the same manifold producing coherent beams parameterized by the pointing angle, with two additional parameters per targeted sidelobe.

Define the initial control pattern for a one-dimensional array scanning to the angle $\theta_0$ as:
\begin{eqnarray}
m(x_n;\theta_0) &=& \frac{\Re(\Psi_{intf}(x_n;\theta_0)) + 1}{2}\\
&=& \frac{\Re(\exp(i (k_s - k_f \sin\theta_0) x_n)) + 1}{2}\\
&=& \frac{\cos((k_s - k_f \sin\theta_0) x_n) + 1}{2}
\label{eq:controlone-dimensional}
\end{eqnarray}

%Figure: measured result
\begin{figure*}[!t]
\centerline{\includegraphics[width=3.5in]{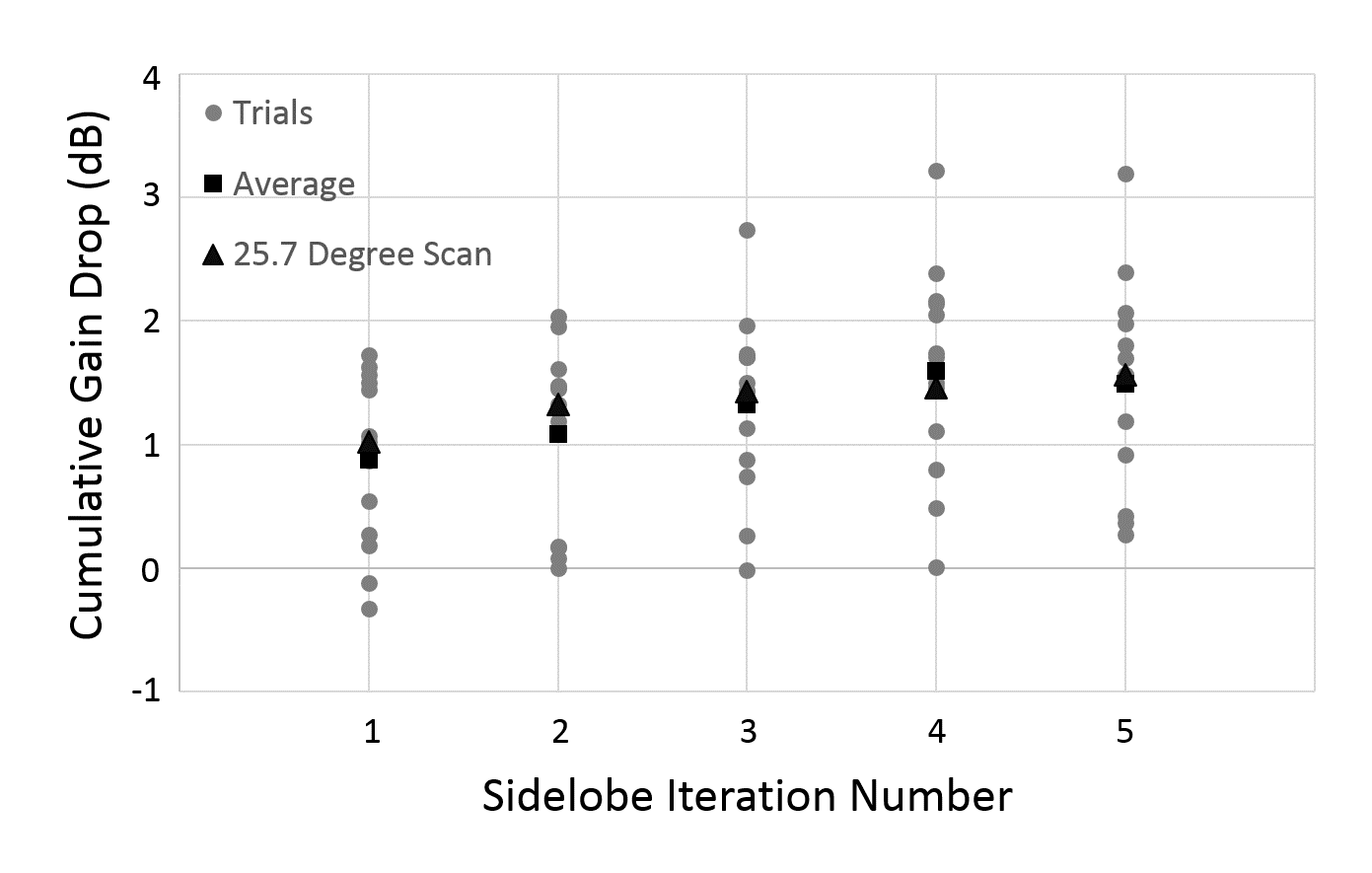}%\label{fig_first_case}}
\hspace{1pt}
\includegraphics[width=3.5in]{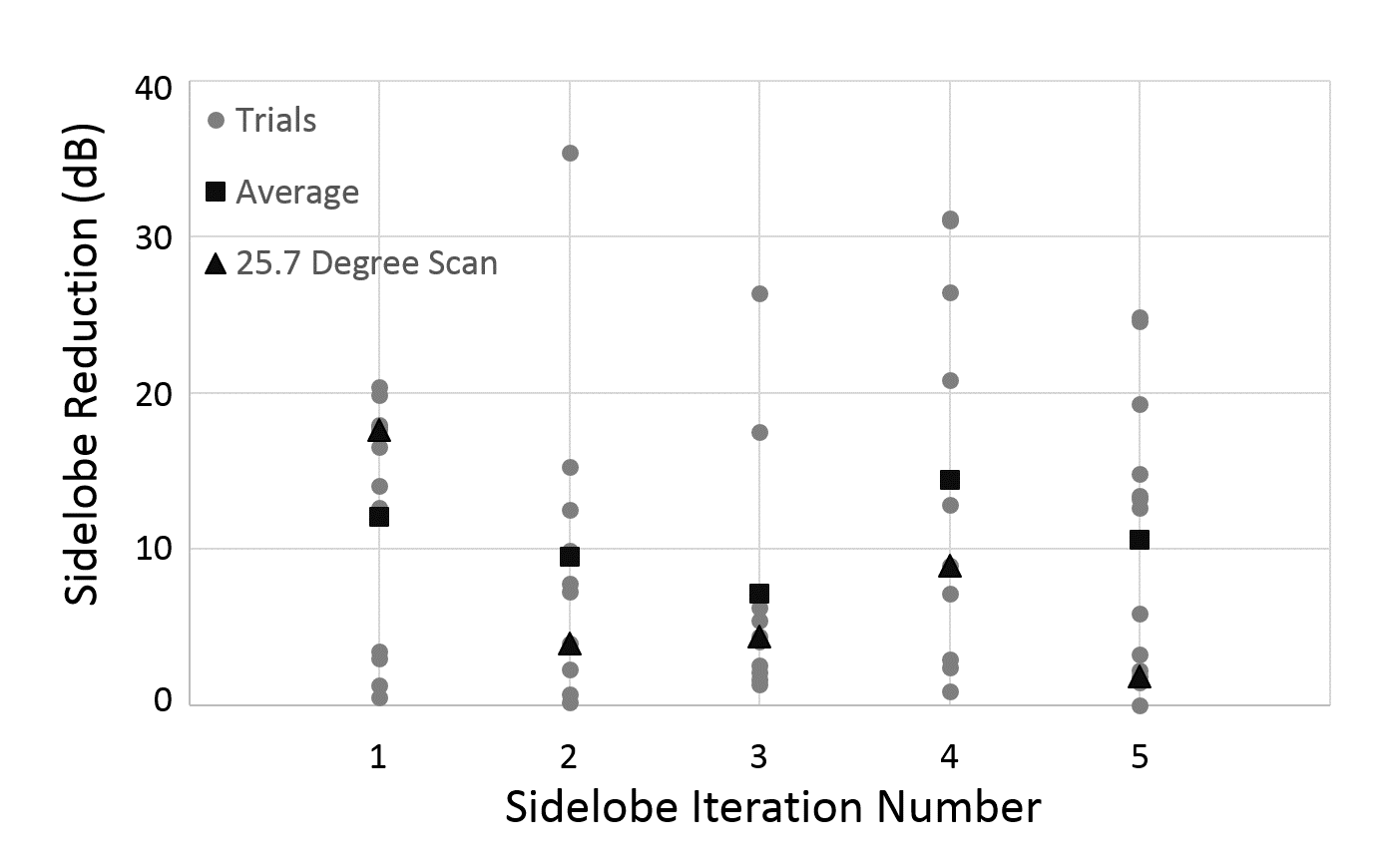}}%
\caption{(a) Cumulative reduction in gain (dB)  and (b) sidelobe reduction (dB) as a function of the number of targeted sidelobes over 13 trials at equally spaced scan angles between $-30^{\circ}$ $\theta$ and $+30^{\circ}$ $\theta$ with the average reduction (square) and the reduction for the scan angle $25.7^{\circ}$ with complete results shown in Figure \ref{fig:FancyImage} (d).}
\label{fig:GainReduction}
\end{figure*}

The algorithm to reduce particularly high-energy sidelobes destructively interferes a second \lq\lq main beam\rq\rq \hspace{1pt} with the targeted sidelobe. In control, it amounts to mangling the original single-tone cosine wave by superimposing the auxiliary control pattern to produce the image of the sidelobe upon it. The effect will be a two-toned (or multi-toned if more than one sidelobe is targeted) control waveform.  

The parameterized control pattern which reduces a sidelobe at a given sidelobe angle $\theta_{1}$ is a renormalization of the weighted sum of the original control pattern and the auxiliary control pattern:
\begin{eqnarray}
m_{sum}(x_n;\theta_0) &=& \cos((k_s - k_f \sin\theta_0) x_n) +  \\
& &\alpha_1 \cos((k_s - k_f \sin\theta_1) x_n + \varphi_1) \notag
\label{eq:summodulations}
\end{eqnarray}
Then $m_{sum}$ is normalized such that $0\leq m_{sum}\leq 1$ for all $x_n$ as before, see steps 2-4 in Figure \ref{fig:Algorithm}.
%\begin{equation}
%m_{sum}(\mathbf{r};\theta_0) = m(\mathbf{r};\theta_0) + \alpha_1 m(\mathbf{r};\theta_{1},\varphi_1)
%\label{eq:summodulations}
%\end{equation}

Note that there are two parameters in addition to the scan angle $\theta_m$ for each of the $M$ auxiliary cosine waveform control patterns that are added to the base modulation pattern to target each of the $M$ sidelobes of interest, which brings the total parameters of the control to $1+3M$ and the parameters which must be optimized to $2M$.  There are two advantages of having a small number of parameters to be optimized.  One is that the optimization can simply be performed much faster. The second is that it gives one a convenient, compact means of storing optimal patterns on the antenna's disk space when the antenna is deployed. 
%It is much more feasible to store $1+3M$ with $M\leq 5$ parameters per desired pointing angle than to have to store the control value for each of $10,000+$ cells.

The two parameters to be optimized are an amplitude, $\alpha_m$, and a phase-offset, $\varphi_m$, of the auxiliary waveform, see steps 5 and 6 of Figure \ref{fig:Algorithm}. The phase-offset must be correctly defined such that the newly produced lobe is $180^{\circ}$ out of phase with the original sidelobe while the amplitude must be selected to sufficiently reduce the sidelobe.  However, both parameters may have a detrimental effect on the main beam and a trade-off in main beam gain and sidelobe reduction may have to be made.

We investigated the parametric topology when maximal reduction of the targeted sidelobe is the optimization criteria. This parameter space is smooth with respect to each parameter independently. That is, fixing one parameter, the sidelobe level varies smoothly with the second parameter. Figure \ref{fig:smooth} shows two representative curves. Left image of Figure \ref{fig:smooth} shows the sidelobe level when the phase-offset of the auxiliary pattern is fixed (at $2^{\circ}$) and the amplitude is varied from 0 to -1; the sidelobe level varies smoothly and has a single minimum.  Similarly, the right plot of the figure shows smooth variation of the sidelobe level when the amplitude of the auxiliary pattern is fixed at $-0.27$ and the phase is varied through all 360 degrees.

The phase and amplitude are well-behaved and could be directly analytically defined if the system were simply a linear array.  However, as noted before there are many opportunities for errors from build tolerances, inexact characterization, inexact modeling, and the application of the controls themselves affect the resultant far-field. So, we use a backtracking line search in conjunction with a gradient descent method to iteratively optimize one parameter at-a-time. This algorithm guarantees reduction of any targeted sidelobe independent of build or modeling and also control errors, learning optimal solutions for individual strips even in the presence of quantization, non-idealities, and complex interactions.
%\begin{figure}[!t]
%\centering
%\includegraphics[width=3.0in]{amplitudesmooth}
%\caption{Sidelobe level as a function of amplitude of auxiliary pattern for the example of an array pointing at -30 degrees and targeting a sidelobe at +4 degrees}
%\label{fig:amplitudesmooth}
%\end{figure}
%
%\begin{figure}[!t]
%\centering
%\includegraphics[width=3.0in]{phasesmooth}
%\caption{Sidelobe level as a function of phase-offset of auxiliary pattern for the example of an array pointing at -30 degrees and targeting a sidelobe at +4 degrees}
%\label{fig:phasesmooth}
%\end{figure}

%Figure: measured result
\begin{figure*}[!t]
%\centerline{\includegraphics[width=3.75in]{threePlots}%\label{fig_first_case}}
%\hspace{1pt}
%\includegraphics[width=3.5in]{threePlotsZoom}}%
\centerline{\includegraphics[width=7in]{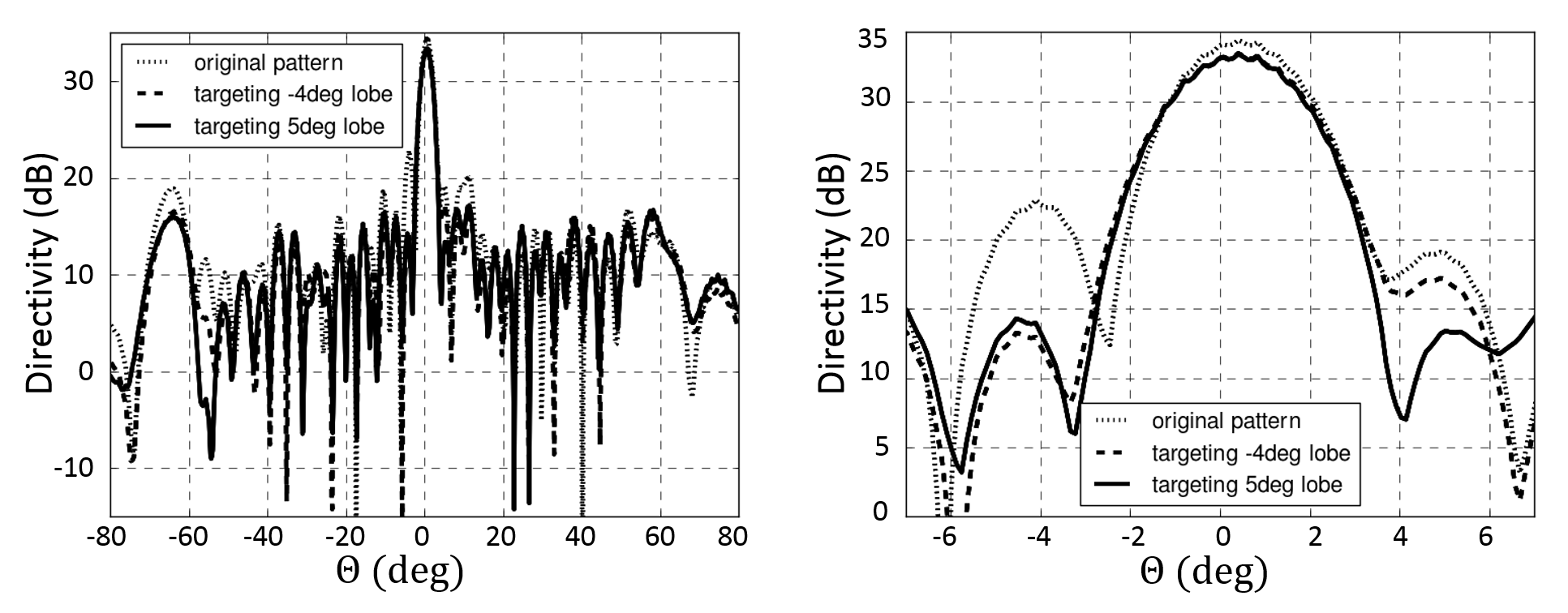}}
\caption{Measured results from a 96 cell single-channel aperture. The aperture was scanned to $0^{\circ}$ $\theta$ and the algorithm was used to reduce the first two sidelobes at $-4^{\circ}$ (by $\approx 8$ dB) and $+5^{\circ}$ degrees (by $\approx 6$ dB). (Left) Entire visible angular region and (right) enlargement of the main beam and targeted sidelobe region.}
\label{fig:MeasuredResults}
\end{figure*}

\section{Results} \label{sec:results}
Results are presented here from using the algorithm of section \ref{sec:algorithm} both in simulation on a DDA modeled single-channel, 260-cell Ka-transmit-band antenna with a hardware taper modeled and from measurement on a small 96 cell single-strip aperture operating in the Ka-transmit-band. These results show the ability of the algorithm to iteratively and significantly reduce undesired sidelobes while not degrading the rest of the far-field pattern, particularly, the main beam gain is acceptably maintained.

Part (c) of figure \ref{fig:FancyImage} shows iterative sidelobe targeting for the 260-cell 30 GHz single-strip antenna simulated with a DDA model. The main beam points to $25.7^{\circ}$ $\theta$, and the algorithm reduces 5 sidelobes at -31.1, -62.0, -64.6, -27.9, and -60.3 degrees, in order. It reduces each of these sidelobes by 17.62, 3.98, 4.43, 8.92, and 1.87 dB, respectively.  The algorithm successfully reduces the original pattern whose first sidelobe was only 8 dB down from the main beam, and produces a pattern where all sidelobes are more than 22 dB down.

Part (c) of figure \ref{fig:FancyImage} also shows a physical representation of what these antenna patterns mean in a mobile use-case.  The antenna is placed in the top of a vehicle, and the arrows show where the primary energies of the antenna are pointing. As the algorithm progresses through sidelobe targeting, the sidelobes shrink rapidly until we are left mostly with just a main beam in the desired direction.  The antenna is no longer at risk of interfering with other satellites' transmissions nor at risk of receiving unacceptable levels of noise making the anticipated transmission impossible to read.

Note that the addition of another discretized wave to the surface of the array will create new sidelobes.  However, the amplitude of these new lobes will be small, since the additional pattern will be added in small \lq\lq doses\rq\rq \hspace{1pt} compared with the original pattern.  It is even sometimes the case that some of these added sidelobes cancel out (or reduce) other pre-existing sidelobes.

The sidelobe-canceled pattern will have reduced gain in the main beam as small amounts of the controlled energy are being redirected to point in the direction of the sidelobe.  However, this reduction will be slight in comparison to the large drop in sidelobe amplitude since the original control waveform will still dominate. The addition of more auxiliary patterns on top of the original waveform will redirect more of the energy away from the main beam.  

Figure \ref{fig:GainReduction} shows the gain reduction as a function of the number of sidelobes killed-off for several trials of the algorithm on the DDA modeled antenna strip. The figure also shows the accompanying reduction in the sidelobe.  The gain drops significantly with the first targeted sidelobe and then degrades more slowly as further sidelobes are targeted.  This is largely due to the fact that the first sidelobe targeted is the largest, and the amplitude $\alpha$ which is optimal to reduce this sidelobe is much larger than the amplitude necessary for a smaller sidelobe.  Thus, with the first, large sidelobe, the auxiliary pattern skews the primary pattern much more heavily than any subsequent addition. Note, however, that the gain reduction is significantly less than the improvement in the sidelobe level.

Figure \ref{fig:MeasuredResults} shows the results of using this algorithm on a real antenna. The antenna is intended to scan to broadside ($0^{\circ}$ $\theta$), and the first two sidelobes, first $-4^{\circ}$ and then $+5^{\circ}$, are reduced with the algorithm. The $-4^{\circ}$ sidelobe was reduced by approximately 8 dB and the $+5^{\circ}$ sidelobe was reduced by approximately 6 dB.  Note that the rest of the pattern is relatively unperturbed, and in some place is noticeably improved even though those sidelobes were not targeted.  Also note that after the $-4^{\circ}$ lobe is targeted, the $+5^{\circ}$ sidelobe correction did not alter, and particularly it did not degrade, the $-4^{\circ}$ sidelobe improvement by much.
Since the modulation patterns are independent, they do not perturb the pattern over-all, only in specific places, those places in their own spectrum where large sidelobes may appear.

\section{Conclusion}

This paper addresses sidelobe cancellation for the optimization of the radiated far-field pattern of a holographic metamaterial antenna. This paper is a first work on the topic:  providing an efficient, robust and adaptive algorithm for optimizing performance of an MSA. 
Considering the control from both holography and phased-array theory gives one a different understanding of the means by which errors may be introduced, and an understanding of both ultimately points to different ways by which to improve the control. 
%In particular, we demonstrated that the controls of the antenna can be usefully understood as the means of recording a holographic picture. 
%Errors in the combination of recorded picture and re-illumination wave will inevitably produce errors in the reproduced three-dimensional image.

After theoretical development, an algorithm to reduce sidelobes was introduced that mitigates some of the errors that are inevitably induced by the finite array size, the pixelation of the array, the quantization of the controls, and the effect the controls themselves have upon the intended reference illuminating wave. The algorithm was successfully demonstrated to significantly reduce sidelobes and maintain main beam gain for both the modeled system and in experiment.

There are several directions to extend this work including exploring methods for full planar aperture optimization.  
%The algorithm could be immediately applied on a full aperture as implemented on the single channel, adding the wave interference control pattern of a sidelobe to all the channels of the aperture simultaneously. However, with more than one row of cells available, one should learn to exploit adjacent rows to cancel sidelobes instead of targeting all sidelobes of a single channel with a single channel.  This could prevent main beam gain degradation. Still other approaches could be taken for optimization of a full-aperture.  
%In particular, the full capabilities of the algorithm presented here should be studied.  Theoretically, this algorithm can be extended to accomplish null-steering that is beneficial for interference removal, produce multiple beams simultaneously that can be used to seamlessly hand-off between satellites, and otherwise achieve desired radiation pattern characteristics. Rigorous study of the robustness of the algorithm to errors in the control due to quantization and specific physical antenna defects (detuning of certain cells, resonance variances with temperature or overtime, etc.) will be necessary for wide-spread deployment of this optimization approach.
%Methods of extending this algorithm to a 
However, perhaps of greatest interest is using this optimization algorithm for real-time control and adaptation in the field, appropriately using modeled predictions and feedback from the antenna's sensors and the satellite communications hub to optimize the pattern on-the-fly. Modeled results along with historical data could be used to provide excellent starting points for the optimization with feedback from the hub allowing for live updating and optimization of the control pattern.

Given the growing importance of mobile satellite technologies, the practical engineering design and performance of a reconfigurable holographic metamaterial antenna is timely for consideration.  
%The MSA is a low-power device that is flat, thin and lightweight and achieves active electronic scanning without any mechanical moving parts.  
We have demonstrated an efficient, software-driven method by which the MSA can operate in a dynamic environment and adaptively suppress the production of detrimental sidelobes and optimize beam steering performance. This first study of its kind for optimizing MSA operation shows that it can be done provided good algorithms are implemented.

\bibliographystyle{ieeetr}
\bibliography{REFS_ALL}
%\begin{thebibliography}{1}
  %0.5em minus 0.4em\relax Harlow, England: Addison-Wesley, 1999.
%
%\end{thebibliography}


\begin{IEEEbiography}[{\includegraphics[width=1in,clip,keepaspectratio]{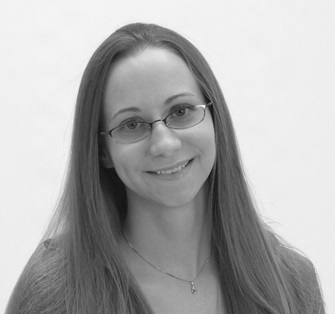}}]{Mikala C. Johnson}
received the B.S. degree in industrial engineering from New Mexico State University, Las Cruces, NM, in 2008.  In 2009, she received her M.S. in applied mathematics from the University of Washington, Seattle, WA, where she is currently working toward the Ph.D. degree in applied mathematics. She is also currently employed as a mathematical research scientist at Kymeta Corporation.
\end{IEEEbiography}

\begin{IEEEbiography}[{\includegraphics[width=1in,clip,keepaspectratio]{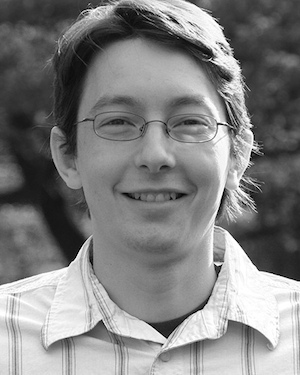}}]{Steven L. Brunton}
received the B.S. degree in mathematics with a minor in control and dynamical systems from the California Institute of Technology, Pasadena, CA, in 2006, and the Ph.D. degree in mechanical and aerospace engineering from Princeton University, Princeton, NJ, in 2012. He is currently an Acting Assistant Professor of applied mathematics at the University of Washington.
\end{IEEEbiography}

%\newpage

\begin{IEEEbiography}[{\includegraphics[width=1in,clip,keepaspectratio]{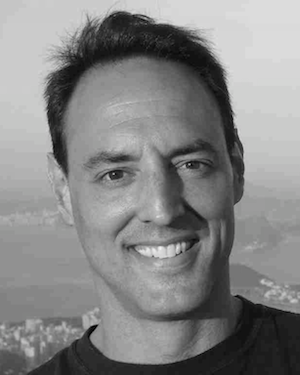}}]{J. Nathan Kutz}
received the B.S. degree in physics and mathematics from the University of Washington, Seattle, WA, in 1990, and the Ph.D. degree in applied mathematics from Northwestern University, Evanston, IL, in 1994. He is currently a Professor and Chair of applied mathematics at the University of Washington.
\end{IEEEbiography}

\begin{IEEEbiography}[{\includegraphics[width=1in,clip,keepaspectratio]{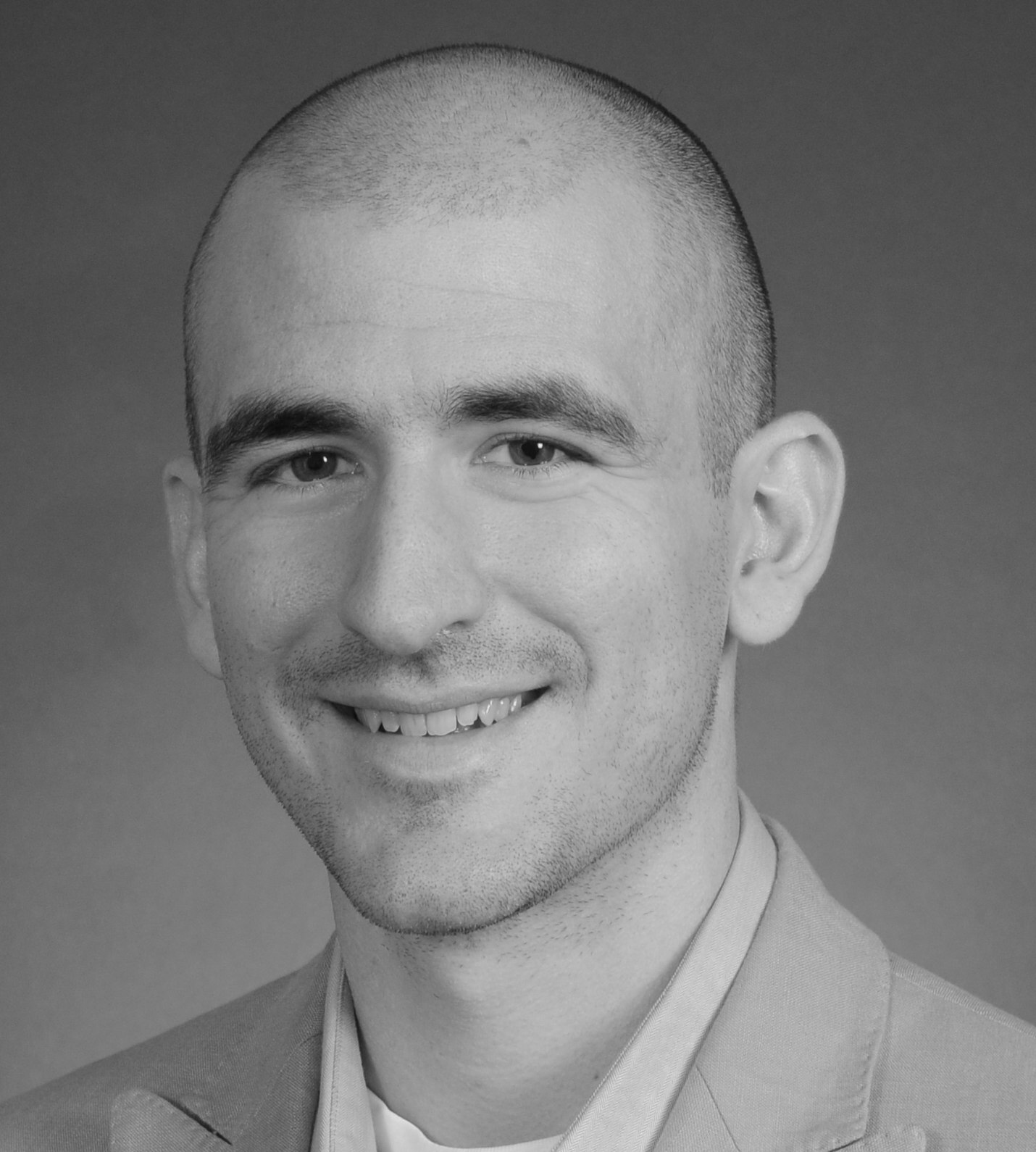}}]{Nathan B. Kundtz}
received the B.S. degree in engineering physics from Case Western Reserve University, in 2004, and the M.S. in electrical engineering and Ph.D. degree in physics from Duke University in 2008 and 2009, respectively. He is a founder, Executive Vice President and Chief Technology Officer of Kymeta Corporation.
\end{IEEEbiography}


%\vfill


\end{document}